\documentclass[fleqn,usenatbib]{mnras}

\usepackage{newtxtext,newtxmath}

\usepackage[T1]{fontenc}

\DeclareRobustCommand{\VAN}[3]{#2}
\let\VANthebibliography\thebibliography
\def\thebibliography{\DeclareRobustCommand{\VAN}[3]{##3}\VANthebibliography}

\usepackage{graphicx}    % Including figure files
\usepackage{amsmath}    % Advanced maths commands

\title[Bayesian Anomaly Detection for Ia Cosmology]{Bayesian Anomaly Detection for Ia Cosmology: Automating SALT3 Data Curation}

\author[Leeney et al.]{
S. A. K. Leeney$^{1,2}$\thanks{E-mail: sakl2@cam.ac.uk},
W. J. Handley$^{2,3}$,
H. T. J. Bevins$^{1,2}$
and E. de Lera Acedo$^{1,2}$
\\
$^{1}$Astrophysics Group, Cavendish Laboratory, University of Cambridge, J. J. Thomson Avenue, Cambridge CB3 0HE, UK\\
$^{2}$Kavli Institute for Cosmology in Cambridge, University of Cambridge, Madingley Road, Cambridge CB3 0HA, UK\\
$^{3}$Institute of Astronomy, University of Cambridge, Madingley Road, Cambridge CB3 0HA, UK
}

\date{Accepted XXX. Received YYY; in original form ZZZ}

\pubyear{\the\year{}}

\begin{document}
\label{firstpage}
\pagerange{\pageref{firstpage}--\pageref{lastpage}}
\maketitle

\begin{abstract}
Precision cosmology with Type Ia supernovae (SNe Ia) requires robust quality control of large, heterogeneous datasets. Current data processing often relies on manual, subjective rejection of photometric data, a practice that is not scalable for forthcoming surveys like the Vera C. Rubin Observatory's Legacy Survey of Space and Time (LSST). We present a Bayesian framework that automates this step by integrating anomaly detection directly into the light curve fitting process. While the framework is model-agnostic and compatible with any spectral energy distribution (SED) model, here we demonstrate its application with the SALT3 model, implemented fully on GPU using \texttt{JAX-bandflux} for computational efficiency. Our method models the probability of each photometric measurement being anomalous with respect to the model, simultaneously inferring the physical supernova parameters and each data point's posterior probability of being a contaminant. Applying this framework to the Hawaii Supernova Flows dataset, we demonstrate its three principal capabilities: (i) robust mitigation of isolated outliers; (ii) automated identification and rejection of entirely corrupted bandpasses; and (iii) preservation of valid data by flagging only specific anomalous points within otherwise usable filters. We also find that contaminants are systematically brighter and bluer, which if uncorrected could bias subsequent cosmological inference. 
\end{abstract}


%%%%%%%%%%%%%%%%%%%%%%%%%%%%%%%%%%%%%%%%%%%%%%%%%%

%%%%%%%%%%%%%%%%% BODY OF PAPER %%%%%%%%%%%%%%%%%%

\section{Introduction}

Type Ia supernovae (SNe Ia) are powerful cosmological probes. Their utility stems from their nature as standardisable candles, allowing for precise distance measurements across vast cosmic scales \citep{Huang2019, Jha2019}. Seminal observations of high redshift SNe Ia provided the first direct evidence for an accelerating expansion of the Universe, a discovery that reshaped modern cosmology and implied the existence of dark energy \citep{Riess1998, Perlmutter1999}.

The contemporary era of supernova cosmology has moved beyond discovery to focus on precision measurements of cosmological parameters \citep{Gall2024, Risaliti2018, Liang2008}. This pursuit requires the compilation and analysis of large, heterogeneous datasets, combining observations from numerous ground based and space based telescopes \citep[e.g.,][]{Scolnic2022}. The inherent diversity in instrumentation, observational conditions, and data reduction pipelines introduces complex systematic uncertainties \citep{Goobar2008, Menard2009, Mandel2016}. The management of these systematic effects has become a central challenge, addressed by a growing suite of sophisticated hierarchical Bayesian models. These frameworks are designed to model the population distributions of supernova and host galaxy properties, allowing for the principled propagation of uncertainties. Applications include disentangling intrinsic colour magnitude relations from dust reddening \citep{Mandel2011, Mandel2014, Mandel2016_dust, Mandel2017}, handling photometric classification and selection biases through frameworks such as BEAMS and BAHAMAS \citep{Hlozek2012, Shariff2015}, and incorporating photometric redshift uncertainties \citep{Roberts2017}. Indeed, ongoing investigations are exploring the impact of cross calibration between different surveys and the choice of light curve modelling frameworks on these results \citep{arXiv:2506.05471, arXiv:2410.13747}. Consequently, the processing of these combined datasets is an intricate and often time consuming task, frequently reliant on significant manual inspection and quality control to ensure the integrity of the final sample \citep{Chotard2011, Hauret2018, Gonzalez2020}.

The significance of addressing these systematic effects has been brought into sharp focus by the recent five year results from the Dark Energy Survey (DESY5). This dataset, one of the largest and most uniform high redshift SNe Ia samples to date, provides powerful new constraints on cosmological parameters. Analysis of the DESY5 sample has revealed intriguing tensions with the standard Lambda Cold Dark Matter ($\Lambda$CDM) model, with some studies interpreting these findings as evidence favouring a dynamic form of dark energy \citep{arXiv:2406.06389, arXiv:2405.03983}. The interpretation of these cosmological implications continues to be debated, emphasising the importance of meticulous scrutiny of the underlying data analysis \citep{arXiv:2503.17342}. The DESY5 findings therefore serve as a powerful contemporary example, underscoring that as statistical power increases, the reliability of cosmological inference becomes increasingly dependent on robust, objective, and reproducible data processing methods.

A particularly challenging step in this process is the selection of photometric data for inclusion in the light curve fit. For a given supernova, analysts often make decisions to accept or reject entire photometric bandpasses based on visual inspection or goodness of fit metrics. This procedure necessarily involves subjective judgements and may contribute to differences between analyses. One consideration with this approach is that an entire filter containing many valid data points may be rejected due to a few outliers or a short period of poor observing conditions, potentially discarding useful information. This problem of data curation is set to become intractable with the advent of next generation surveys. The Vera C. Rubin Observatory's Legacy Survey of Space and Time (LSST) is expected to discover millions of SNe Ia, rendering any form of manual vetting or subjective filter selection entirely infeasible \citep{Ivezic2019, LSST_DESC2012, Mandelbaum2019}. The anticipated scale of LSST's supernova sample will require automated methods for identifying and mitigating systematic effects \citep{Arendse2023, Wojtak2019, Hlozek2019, Foley2018}.

These escalating challenges in data volume and complexity necessitate a fundamental shift away from manual interventions towards fully automated, statistically robust, and reproducible methods for data processing. Hierarchical Bayesian models, such as BayeSN \citep{Mandel2020, Thorp2021, Dhawan2022}, have demonstrated the potential for improved precision in SNe Ia distance measurements by simultaneously modelling spectral energy distributions and dust extinction. Beyond simply identifying contaminated data, understanding how anomalous measurements systematically bias physical parameters is crucial for precision cosmology. The colour parameter in particular, which captures both intrinsic supernova variation and dust extinction, directly impacts distance moduli through the Tripp relation and can introduce systematic errors if contamination effects are not properly quantified. It would be beneficial to have a unified framework that can simultaneously fit the physical supernova model, account for potential data contamination in a principled manner, and quantify the systematic biases introduced by anomalous data. In this paper, we present such an approach: a Bayesian anomaly detection framework designed for the fully automated processing of SNe Ia light curves with integrated contamination quantification.

This Bayesian anomaly detection framework, adapted from techniques developed for 21cm cosmology~\citep{Leeney2022, Anstey2023, Roque2023, de2022reach}, directly integrates anomaly detection into the likelihood function of the light curve fitting process. Our implementation leverages GPU acceleration to achieve the computational efficiency necessary for processing the vast data volumes anticipated from next generation surveys. With LSST expected to discover 3 to 4 million supernovae over its 10 year survey and generate over 10 million transient alerts nightly \citep{ivezic2019lsst}, traditional CPU based processing pipelines will struggle to keep pace with this data avalanche. Our GPU accelerated framework demonstrates that modern parallel computing architectures can enable the real time, automated analysis required for the LSST era.

The anomaly detection component is agnostic to the underlying physical model, allowing it to be coupled with any light curve fitter such as BayeSN or SNooPy; for this work, we integrate it with the widely used SALT3 model \citep{Guy2005orig, Guy2007, Kenworthy2021}. This approach extends traditional data curation methods in three important ways. Firstly, it robustly mitigates the impact of isolated, anomalous data points that can otherwise bias parameter estimates. Secondly, it automates the process of filter selection, providing a probabilistic assessment of whether a given bandpass is reliable without human intervention. Thirdly, and most significantly, our framework facilitates enhanced data preservation. By identifying and modelling only the specific anomalous epochs within a given filter, it retains the surrounding valid data points that would otherwise be lost when comprehensive filter level rejection is necessary, as demonstrated in our application to the carefully processed Hawaii Supernova Flows (HSF) \citep{do2025hawaii} dataset. This ensures that the maximum amount of information is extracted from the data, leading to more precise and robust cosmological constraints than is possible with all-or-nothing filter rejection schemes.

The remainder of this paper is structured as follows. In Section~\ref{sec:theory}, we present the theoretical foundation of our Bayesian anomaly detection framework, including the mathematical formulation of the model and its integration with the SALT light curve template. Section~\ref{sec:methods} describes the practical implementation details, including the computational approach using GPU accelerated model evaluation and the application to the Hawaii Supernova Flows dataset. In Section~\ref{sec:results}, we demonstrate the framework's performance through both a comprehensive statistical comparison across the full dataset and detailed case studies of individual supernovae that illustrate the three key benefits of the method. Finally, Section~\ref{sec:conclusion} summarises our findings and discusses the broader implications for precision cosmology with future large scale surveys.

\section{Theory}
\label{sec:theory}

In this section, we outline the theoretical underpinnings of our automated data processing framework. We begin with a brief overview of Bayesian inference, before detailing the specific formulation of our Bayesian anomaly detection model. We conclude by introducing the Spectral Adaptive Lightcurve Template (SALT) model, which serves as the physical basis for our analysis.

\subsection{Bayesian Inference}

Bayesian inference offers a principled framework for estimating model parameters
and comparing competing hypotheses \citep{MacKay2003}. For a given model $\mathcal{M}$ with
parameters $\theta$ and a dataset $\mathcal{D}$, Bayes' theorem relates the
posterior probability of the parameters, $\mathcal{P}(\theta) \equiv P(\theta
| \mathcal{D}, \mathcal{M})$, to the likelihood of the data,
$\mathcal{L}(\theta) \equiv P(\mathcal{D} | \theta, \mathcal{M})$, and the prior
probability of the parameters, $\pi(\theta) \equiv P(\theta | \mathcal{M})$. The
theorem is expressed as: \begin{equation} P(\theta | \mathcal{D}, \mathcal{M})
= \frac{P(\mathcal{D} | \theta, \mathcal{M}) P(\theta
| \mathcal{M})}{P(\mathcal{D} | \mathcal{M})} \quad \iff \quad
\mathcal{P}(\theta) = \frac{\mathcal{L}(\theta) \pi(\theta)}{\mathcal{Z}}
\end{equation} The denominator, $\mathcal{Z} \equiv P(\mathcal{D}
| \mathcal{M})$, is the Bayesian evidence. It is obtained by integrating the
product of the likelihood and the prior over the entire parameter space. The
evidence serves as a normalisation constant for the posterior in parameter
estimation problems. However, its value is paramount for model comparison, as it
represents the probability of the data given the model as a whole, naturally
penalising overly complex models. In most real world applications, the posterior
distribution is too complex for an analytical solution and must be evaluated
using numerical methods such as Nested Sampling \citep{Skilling2006, buchner2023nested, ashton2022nested} or other Markov Chain Monte Carlo techniques.

\subsection{Bayesian Anomaly Detection}
\label{subsec:anomaly_detection}

The Bayesian anomaly detection methodology presented here is adapted from a framework originally developed for mitigating radio frequency interference in 21cm cosmology \citep{Leeney2022, Anstey2023, Roque2023, de2022reach}. To account for anomalous data within a Bayesian framework, we formulate the 
problem as a model selection task at the level of individual data points. A 
standard likelihood, which assumes all data points are drawn from a single 
physical model, cannot handle anomalous data points. To address this, we reformulate the 
likelihood to explicitly model the possibility that any data point may be an 
anomaly. For each of the $N$ data points in our dataset, we introduce a binary 
latent variable $\varepsilon_i \in \{0, 1\}$. We define $\varepsilon_i=1$ to 
indicate the data point is "expected" and described by our physical model, and 
$\varepsilon_i=0$ to indicate it is an "anomaly". The 
full set of these variables forms a Boolean mask vector $\varepsilon$. Each photometric data point has its own independent probability $P(\varepsilon_i=0)$, calculated separately for each filter-epoch combination, as observations in different filters are taken at different times and thus have independent contamination probabilities.

The likelihood for the entire dataset $\mathcal{D}$, conditioned on a specific mask $\varepsilon$ and model parameters $\theta$, is the product of the likelihoods for each individual data point:
\begin{equation}
    P(\mathcal{D} | \theta, \varepsilon) = \prod_{i=1}^{N} \left[\mathcal{L}_i(\theta)\right]^{\varepsilon_i} \left[\frac{1}{\Delta_i}\right]^{1-\varepsilon_i}
\end{equation}
Here, $\mathcal{L}_i(\theta)$ is the likelihood for the $i$-th data point being "good", typically a distribution (such as a Gaussian) centred on the physical model's prediction. $\frac{1}{\Delta_i}$ is the likelihood for the point being an anomaly, which we model as a broad, uninformative uniform distribution over a plausible data range $\Delta_{\mathrm{range},i}$, such that $\frac{1}{\Delta_i} = \frac{1}{\Delta_{\mathrm{range},i}}$.

We do not know the true mask $\varepsilon$ a priori. We therefore infer it by ascribing a Bernoulli prior probability to each $\varepsilon_i$, governed by a single hyperparameter $p$, which represents our prior belief that any given data point is an anomaly:
\begin{equation}
    P(\varepsilon_i | p) = p^{1-\varepsilon_i} (1-p)^{\varepsilon_i}
\end{equation}
Assuming each data point's status is independent, the joint probability of the data and the mask is the product of the likelihood and the prior:
\begin{equation}
    P(\mathcal{D}, \varepsilon | \theta, p) = \prod_{i=1}^{N} \left[ \mathcal{L}_i(\theta)(1-p) \right]^{\varepsilon_i} \left[ \frac{p}{\Delta_i} \right]^{1-\varepsilon_i}
\end{equation}
To obtain a likelihood that is independent of the unknown mask $\varepsilon$, we must marginalise over all $2^N$ possible states of the mask vector:
\begin{equation}
    P(\mathcal{D} | \theta, p) = \sum_{\varepsilon \in \{0,1\}^N} P(\mathcal{D}, \varepsilon | \theta, p)
\end{equation}
Since the data points are independent, this sum can be brought inside the product, yielding the exact marginalised likelihood:
\begin{equation}
\label{eq:marginalised_likelihood}
    \mathcal{L}(\mathcal{D}|\theta, p) = \prod_{i=1}^{N} \left( \mathcal{L}_i(\theta)(1-p) + \frac{p}{\Delta_i} \right)
\end{equation}
For large $N$, the direct evaluation of this expression is not possible. However, for a well specified model, the sum in Equation 5 is typically dominated by a single mask, $\varepsilon^\mathrm{max}$, which maximises the joint probability $P(\mathcal{D}, \varepsilon | \theta, p)$. This "dominant mask approximation" is valid provided that the probability of the most likely mask is significantly greater than that of the next most likely mask, which is typically $\varepsilon^\mathrm{max}$ with a single bit flipped:
\begin{equation}
    P(\mathcal{D}, \varepsilon^\mathrm{max} | \theta, p) \gg \max_{j} P(\mathcal{D}, \varepsilon^{(j)} | \theta, p)
\end{equation}
Under this approximation, the log likelihood becomes a sum over the contributions from each data point, where for each photometric observation we choose the more probable of the `expected' or `anomaly' hypotheses:
\begin{equation}
    \log \mathcal{L}(\mathcal{D}|\theta, p) \approx \sum_{i=1}^{N} \max \left\{ \log\mathcal{L}_i(\theta) + \log(1-p), \log p - \log\Delta_i \right\}
\end{equation}
This can be written explicitly as:
\begin{equation}
\begin{split}
    \log \mathcal{L}(\mathcal{D}|\theta, p) &= \sum_{i=1}^{N} \begin{cases}
        \log\mathcal{L}_i(\theta) + \log(1-p), & \text{if } \log\mathcal{L}_i(\theta) + \log(1-p) \\
        & \quad > \log p - \log\Delta_i \\
        \log p - \log\Delta_i, & \text{otherwise}
    \end{cases}
\end{split}
\end{equation}
This establishes a dynamic, model aware classification for each data point. A point is flagged as an anomaly if the likelihood of it being so, penalised by the prior probability $p$, exceeds the likelihood of it conforming to the physical model. Rearranging the condition for a point being classified as "good" reveals a connection to the logit function:
\begin{equation}
    \log\mathcal{L}_i(\theta) + \log\Delta_i > \log\left(\frac{p}{1-p}\right) \equiv \mathrm{logit}(p)
\end{equation}
This formulation effectively places a floor on the log likelihood for each data point, preventing single outliers from dominating the total likelihood and biasing parameter estimates. The hyperparameter $p$ can be treated as a free parameter and inferred from the data, allowing the framework to learn the intrinsic data quality automatically.

\subsection{The SALT Model}
\label{subsec:salt_model}

The physical model underlying the "good" likelihood, $\mathcal{L}_i(\theta)$, is provided by the Spectral Adaptive Lightcurve Template (SALT) framework \citep{Guy2005, Guy2007}. SALT is a phenomenological model describing the spectral energy distribution (SED) of a Type Ia supernova as a function of phase and wavelength. The most recent public version, SALT3, benefits from a larger and more diverse training set \citep{Kenworthy2021}. The SALT model has been widely adopted in cosmological analyses and forms the foundation for modern SNe Ia distance measurements \citep{Scolnic2022, Hayes2024}.

The SALT model describes the spectral flux, $F(p, \lambda)$, at phase $p$ and rest frame wavelength $\lambda$ with a set of parameters $\theta = \{x_0, x_1, c\}$. These correspond to the overall amplitude ($x_0$), a light curve shape or "stretch" parameter ($x_1$), and a colour parameter ($c$). The flux is constructed from a mean spectral surface, $M_0$, a first order variation surface, $M_1$, and a colour law, $CL(\lambda)$:
\begin{equation}
\label{eq:salt_flux}
    F(p, \lambda) = x_0 \times [M_0(p, \lambda) + x_1 M_1(p, \lambda)] \times \exp(c \times CL(\lambda))
\end{equation}
The colour parameter $c$ is particularly important as it captures both intrinsic colour variation of the supernova and dust extinction along the line of sight. Positive $c$ values indicate redder supernovae (more extinction or intrinsically redder), while negative values indicate bluer supernovae. The colour term modifies the spectrum exponentially through $\exp(c \times CL(\lambda))$, where $CL(\lambda)$ typically increases toward bluer wavelengths.

For a given photometric observation $d_i$ with uncertainty $\sigma_i$ in a specific bandpass, the model prediction is found by integrating this SED over the instrumental transmission profile. The predicted bandpass integrated flux is given by:
\begin{equation}
    f_i(\theta) = \int_{\lambda_{\mathrm{min}}}^{\lambda_{\mathrm{max}}} F(p, \lambda) \cdot T(\lambda) \cdot \frac{\lambda}{hc} \, d\lambda
\end{equation}
where $T(\lambda)$ is the transmission function of the instrumental bandpass, and $\frac{\lambda}{hc}$ converts from energy flux to photon flux. The "good" likelihood is then given by $\mathcal{L}_i(\theta) = \mathcal{N}(d_i | f_i(\theta), \sigma_i^2)$, where $f_i(\theta)$ is the predicted bandpass integrated flux. The goal of our framework is to infer the posterior distributions for the supernova parameters $\theta$ and the anomaly hyperparameter $p$.

\section{METHODS}
\label{sec:methods}

The Bayesian anomaly detection framework described in Section~\ref{sec:theory} is designed for modularity. The anomaly-aware likelihood modification is independent of the specific supernova model used for the light curve evaluation. While this paper implements the framework with SALT3, replacing it with an alternative model would only require substituting the function call that evaluates the model flux for a given set of parameters.

The theoretical framework described in Section~\ref{sec:theory} provides a general approach for Bayesian anomaly detection. Here, we detail its practical implementation for the specific application of processing multi bandpass photometric data from Type Ia supernovae. This involves defining the full likelihood for the dataset, outlining the computational methods for evaluating the physical model, and describing the numerical techniques used to explore the resulting posterior distribution.

Our dataset for a single supernova consists of multiple photometric flux
measurements, $\{d_{ij}\}$, taken at different times and across several distinct
bandpasses, indexed by $j$. Each measurement has an associated uncertainty,
$\sigma_{ij}$. The full likelihood for the entire dataset, $\mathcal{D}$, given
the SALT parameters $\theta$ and the anomaly hyperparameter $p$, is constructed
by applying the marginalised likelihood from Equation~\eqref{eq:marginalised_likelihood}. We take the product
over all bandpasses and all data points within each band: \begin{equation}
\mathcal{L}(\mathcal{D}|\theta, p) = \prod_{j}^{\mathrm{bands}}
\prod_{i}^{\mathrm{obs}} \left( \mathcal{L}_{ij}(\theta)(1-p)
+ \frac{p}{\Delta_{ij}} \right) \end{equation} The "good" likelihood for each point,
$\mathcal{L}_{ij}(\theta)$, is a Gaussian distribution centred on the SALT model prediction:
\begin{equation}
\mathcal{L}_{ij}(\theta) = \frac{1}{\sqrt{2\pi\sigma_{ij}^2}} \exp\left(-\frac{(d_{ij} - f_{ij}(\theta))^2}{2\sigma_{ij}^2}\right)
\end{equation}
where $f_{ij}(\theta)$ is the model flux predicted by SALT for the given parameters $\theta$, $d_{ij}$ is the observed flux, and $\sigma_{ij}$ is the flux uncertainty. The "bad" model, $\frac{p}{\Delta_{ij}}$, is a broad uniform distribution covering the plausible range of flux values for that observation. The log likelihood therefore becomes:
\begin{equation}
\begin{split}
\log \mathcal{L}(\mathcal{D}|\theta, p) &\approx \sum_{j}^{\mathrm{bands}} \sum_{i}^{\mathrm{obs}} \max \left\{ \log\mathcal{L}_{ij}(\theta) + \log(1-p), \right.\\
&\left. \log p - \log\Delta_{ij} \right\}
\end{split}
\end{equation}

A significant computational challenge lies in the evaluation of the model flux, $f_{ij}(\theta)$. The SALT model, as described by Equation~\eqref{eq:salt_flux}, provides a spectral energy distribution (SED) as a function of phase and wavelength. To compare this with a photometric observation, the model SED must be integrated over the transmission profile of the corresponding instrumental bandpass. This integration is a computationally intensive operation that must be performed for every data point at every step of the posterior sampling process, which can make comprehensive Bayesian analyses computationally challenging for large datasets.

To overcome this bottleneck, we use the \texttt{JAX-bandflux} library \citep{leeney2025jax}. This package provides a fully differentiable implementation of the SALT3 model and bandpass integration routines, built upon the JAX framework for high performance numerical computing. By leveraging JAX, the entire model evaluation can be just in time compiled and executed on Graphics Processing Units (GPUs). This enables vectorisation of the bandflux calculation components, which significantly accelerates the computation. This computational efficiency is critical for making the analysis of large datasets tractable and enables the use of GPU-accelerated sampling techniques. While this work leverages GPUs via JAX for maximum performance on large datasets, the anomaly detection framework itself is hardware-agnostic and can be run on standard CPUs, making it accessible for a wide range of analysis scales.

With the likelihood and its efficient computation defined, the final step is to explore the multi dimensional posterior distribution $P(\theta, p | \mathcal{D})$ to infer the model parameters. The complexity of the posterior landscape makes an analytical solution impossible, necessitating the use of a numerical sampling algorithm. To initialise the sampling process, we first obtain a maximum likelihood estimate using the Nelder-Mead simplex method \citep{Nelder1965}, which provides a robust starting point for the posterior exploration without requiring gradient information. For this work, we use GPU accelerated Nested Sampling \citep{yallup2025nested}. Nested Sampling explores the parameter space by iteratively moving through nested contours of constant likelihood. This process yields a set of weighted samples from the posterior distribution, which can be used to estimate parameters and their uncertainties. Nested Sampling also calculates the Bayesian evidence, $\mathcal{Z}$, as a primary output, which can be used for model comparison. It is important to note, however, that our framework is sampler agnostic. The core methodology of the anomaly aware likelihood is independent of the technique used to explore it, and other methods such as other Markov Chain Monte Carlo techniques could be used.

\subsection{Dataset}

To test and validate our framework, we use data from the Hawaii Supernova Flows (HSF) project \citep{do2025hawaii}. The photometric data for each event are heterogeneous, comprising optical light curves from public all sky surveys such as ATLAS \citep{Tonry2018}, ASAS-SN \citep{Shappee2014}, and ZTF \citep{Bellm2019}, combined with dedicated near infrared follow up photometry from UKIRT \citep{Lawrence2007}, primarily in the J band. The diversity of these data sources presents significant data quality challenges, making this dataset an ideal testbed for automated methods. The comprehensive HSF analysis employs careful manual inspection of light curves combined with statistical criteria to ensure data quality. To enable a systematic comparison with our automated framework, we develop a reproducible rule that approximates traditional filter selection approaches: we test all possible combinations of filters, calculate the $\chi^2$ goodness-of-fit for each combination, and select the combination yielding the lowest $\chi^2$ while requiring at least three distinct photometric bands.

\subsection{Analysis Cases}
\label{subsec:analysis_cases}

We assess the performance of our Bayesian anomaly detection framework by comparing it against two alternative analysis schemes. This allows us to quantify the improvement over both a naive approach and the current standard practice. The three distinct analysis cases are as follows.

\textit{Case A} represents a naive fit, in which a standard SALT analysis is performed using all available photometric data from all bandpasses. No data cleaning, outlier rejection, or anomaly detection is applied. This case serves as a baseline to demonstrate the impact of unmitigated data contamination on the resulting parameter estimates. \textit{Case B} represents a systematic approximation of traditional filter selection methods, inspired by the careful curation performed in \citet{do2025hawaii}. We employ the reproducible rule described above: testing all possible filter combinations and selecting the one with the best $\chi^2$ fit while requiring at least three distinct photometric bands. This approach is designed to approximate the outcome of traditional expert analyses with a systematic and reproducible rule. This provides a necessary and consistent baseline against which the performance of our fully automated framework can be quantitatively assessed. \textit{Case C} is our proposed method, where the full Bayesian anomaly detection framework is applied. The analysis uses all available photometric data, just as in \textit{Case A}, but incorporates the anomaly model described in Section~\ref{subsec:anomaly_detection}. This allows the model to probabilistically identify and downweigh individual anomalous data points in a fully automated and statistically principled manner, without discarding any data a priori.

The outputs from the Nested Sampling runs for each analysis case are processed and visualised using the \texttt{anesthetic} package \citep{Handley2019}. We generate posterior probability distributions for the key supernova parameters to quantitatively compare the constraints from each method. We also produce light curve plots showing the data with the best fitting model from each case overlaid, providing a direct visual comparison of their performance.

\subsection{Similarity Analysis}

To validate our automated framework, we perform a similarity analysis to quantitatively compare its results to our systematic filter selection approach (Case B). This analysis tests whether the two methods produce statistically consistent parameter estimates and examines any systematic differences that emerge, particularly in cases where the automated framework preserves data that would otherwise be discarded. For each supernova in our sample, we obtained posterior samples for the four key SALT3 parameters $\boldsymbol{\theta} = \{t_0, \log x_0, x_1, c\}$, where $t_0$ is the time of maximum light, $\log x_0$ is the logarithm of the normalisation parameter, $x_1$ is the light curve shape parameter, and $c$ is the colour parameter. The systematic differences observed, particularly in the colour parameter, are explained through the contamination quantification analysis described in Sections~\ref{subsec:contamination_quantification} and \ref{subsec:population_contamination}.

For each parameter $\theta_i$ and each supernova $j$, we computed the normalised difference between the two methods (Case B and Case C):
\begin{equation}
\Delta\theta_{i,j} = \frac{|\mu_{i,j}^{\mathrm{HSF}} - \mu_{i,j}^{\mathrm{anomaly}}|}{\sqrt{(\sigma_{i,j}^{\mathrm{HSF}})^2 + (\sigma_{i,j}^{\mathrm{anomaly}})^2}}
\end{equation}
where $\mu_{i,j}^{\mathrm{method}}$ is the posterior mean of parameter $i$ for supernova $j$ using the specified method, and $\sigma_{i,j}^{\mathrm{method}}$ is the corresponding posterior standard deviation. This normalised difference represents the parameter difference in units of combined standard deviations, with values $\Delta\theta_{i,j} < 1$ indicating that the methods agree within their combined uncertainty.

\subsection{Contamination Quantification}
\label{subsec:contamination_quantification}

Beyond identifying anomalous data points (flux measurements), quantifying their systematic impact on parameter estimation is important for understanding potential biases in cosmological analyses. Importantly, our framework does not make binary classifications of anomalies; instead, each data point has a continuous posterior probability of being anomalous. We define several metrics that incorporate this probabilistic nature to characterise how contamination affects the inferred SALT3 parameters.

For each data point $i$, the posterior probability of being anomalous is:
\begin{equation}
    P(\varepsilon_i = 0 | \mathcal{D}, \theta) = \frac{p/\Delta_i}{\mathcal{L}_i(\theta)(1-p) + p/\Delta_i}
\end{equation}
This probability is evaluated at each sample from the posterior distribution $P(\theta|\mathcal{D})$, obtained through a numerical sampling procedure. The standardised residuals for each observation are:
\begin{equation}
    r_i = \frac{d_i - f_i(\theta)}{\sigma_i}
\end{equation}
where $d_i$ is the observed flux and $f_i(\theta)$ is the SALT3 model prediction.

To quantify the overall flux bias introduced by anomalous points, we define the brightness contamination:
\begin{equation}
    C_{\mathrm{bright}} = \frac{\sum_i P(\varepsilon_i = 0) \cdot r_i}{\sum_i P(\varepsilon_i = 0)}
\end{equation}
This metric incorporates the probabilistic nature of anomaly detection: each data point contributes to the contamination score weighted by its probability of being anomalous, rather than through binary classification. The resulting weighted average reveals whether the net systematic effect of probable anomalies is to push flux measurements upward (positive) or downward (negative).

The wavelength dependent bias is captured by the colour contamination metric. To quantify whether anomalies preferentially affect certain wavelengths, we compute separate brightness contamination metrics for blue and red portions of the spectrum. These are calculated using the same probability-weighted average as $C_{\mathrm{bright}}$, but only including data points from blue ($\lambda < 5000$ Å, typically ZTF g, ATLAS c) or red ($\lambda > 6000$ Å, typically ZTF r, i, UKIRT J) bands. The colour contamination is then defined as the difference between red and blue contamination:
\begin{equation}
    C_{\mathrm{colour}} = C_{\mathrm{red}} - C_{\mathrm{blue}}
\end{equation}
where
\begin{equation}
    C_{\mathrm{blue}} = \frac{\sum_{i \in \text{blue}} P(\varepsilon_i = 0) \cdot r_i}{\sum_{i \in \text{blue}} P(\varepsilon_i = 0)}, \quad C_{\mathrm{red}} = \frac{\sum_{i \in \text{red}} P(\varepsilon_i = 0) \cdot r_i}{\sum_{i \in \text{red}} P(\varepsilon_i = 0)}
\end{equation}
Positive $C_{\mathrm{colour}}$ indicates that anomalous points preferentially affect red bands, potentially biasing the colour parameter $c$ upward.

These metrics are computed across all posterior samples, propagating uncertainty through the full analysis.

\subsection{Population-Level Contamination Analysis}
\label{subsec:population_contamination}

To assess systematic contamination patterns across the full supernova sample, we compute the contamination metrics for each supernova and examine their distribution and implications. These population level metrics retain the probabilistic nature of our anomaly detection: they represent the probability weighted net effects across all data points, not binary counts of anomalies. The brightness contamination $C_{\mathrm{bright}}$ quantifies whether the probability weighted contribution of potentially anomalous points systematically pushes flux measurements upward (positive values) or downward (negative values). Similarly, the colour contamination $C_{\mathrm{colour}}$ reveals wavelength dependent biases, with negative values indicating that the probability weighted effect preferentially affects blue bands and positive values indicating red band preference.

For the population of supernovae, we examine the distribution of these contamination metrics to identify systematic patterns. The mean and standard deviation across the sample reveal whether contamination effects are consistent or vary significantly between objects. Large standard deviations would indicate heterogeneous contamination requiring object by object treatment, while consistent biases might suggest systematic instrumental or processing effects.

These contamination metrics directly explain the systematic differences observed in SALT3 parameters between our anomaly detection framework and traditional methods. In particular, the colour contamination $C_{\mathrm{colour}}$ provides insight into why the colour parameter $c$ shows larger differences in the similarity analysis: systematic wavelength dependent contamination biases the inferred extinction. Understanding these biases is crucial for future cosmological analyses where they would propagate through the Tripp relation: $\mu = m_B - M + \alpha x_1 - \beta c$, directly affecting distance moduli.

\section{RESULTS}
\label{sec:results}

We have applied our Bayesian anomaly detection framework to all suitable Type Ia supernovae from the Hawaii Supernova Flows dataset \citep{do2025hawaii}. In this section, we first present a global statistical comparison between our method and the systematic approximation of traditional filter selection, before using three representative supernovae to illustrate the framework's different benefits.

\subsection{Similarity Analysis}

Our global similarity analysis confirms that the automated framework (Case C; Section~\ref{subsec:analysis_cases}) produces parameter estimates that are statistically consistent with our systematic filter selection approach (Case B; Section~\ref{subsec:analysis_cases}). The overall agreement demonstrates the reliability of our automated approach. To investigate the systematic differences observed, particularly in the colour parameter, we performed a contamination analysis (Section~\ref{subsec:contamination_analysis}) that establishes how anomalous data points correlate with both brightness and wavelength, revealing systematic patterns in the data quality that can affect parameter estimation.

\begin{figure*}
\centering
\includegraphics[width=0.8\textwidth]{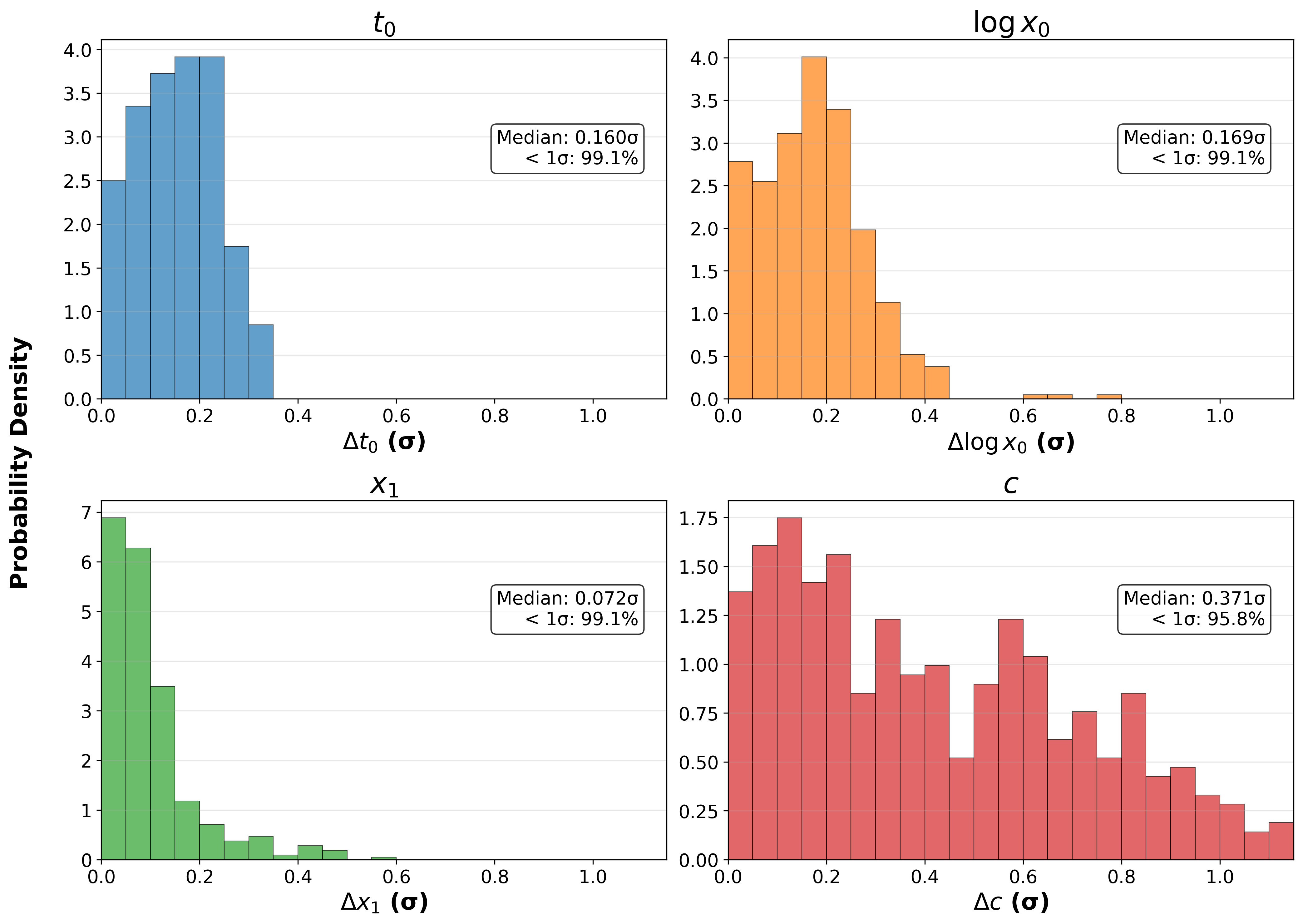}
\caption{A histogram visualising the distributions of the normalised difference, $|\Delta\theta|$, for the four SALT3 parameters, comparing our anomaly detection method to the systematic filter selection approach (Case B) across all supernovae. The vast majority of the probability mass (98.2 per cent) lies well within the 1$\sigma$ similarity threshold, demonstrating that the two methods agree within their combined uncertainties. For the color parameter, the distribution is slightly more spread out with a median of 0.371$\sigma$ and 95.8 per cent of the probability mass within 1 sigma of the traditional analysis.}
\label{fig:similarity_corner}
\end{figure*}

The results, summarised in Figure~\ref{fig:similarity_corner}, show that 98.2 per cent of all parameter estimates agree to within 1$\sigma$ of their combined uncertainty. On an object-by-object basis, 93.9 per cent of supernovae have all four of their parameters agreeing within this 1$\sigma$ threshold. The mean normalised differences are $\Delta t_0 = 0.154\sigma$, $\Delta \log x_0 = 0.175\sigma$, $\Delta x_1 = 0.097\sigma$, and $\Delta c = 0.418\sigma$.

Notably, the colour parameter $c$ (one of the SALT parameters described in Section~\ref{subsec:salt_model}), exhibits a modestly larger average difference ($\Delta c = 0.418\sigma$) than other parameters. This is interesting because it indicates a systematic difference in our method's predictions versus traditional methods' predictions. This suggests that in addition to automating the data curation process, we may be removing anomalies that are systematically influencing subsequent cosmological analyses, as quantified in our contamination analysis (Section~\ref{subsec:contamination_analysis}). By preserving valid data points in partially contaminated filters, a capability we term `filter preservation', the method retains colour information that is lost when entire bandpasses are discarded. This improvement in data retention, while small for individual objects, has the potential to significantly impact cosmological analyses of large samples by reducing biases introduced by coarse data rejection.

\subsection{Contamination Analysis}
\label{subsec:contamination_analysis}

To understand the systematic differences observed in the similarity analysis, particularly for the colour parameter, we performed a contamination analysis across the supernova sample. This analysis quantifies how anomalous data points systematically bias SALT3 parameter estimation through the contamination metrics defined in Section~\ref{subsec:contamination_quantification}.

\begin{figure*}
\centering
\includegraphics[width=0.8\textwidth]{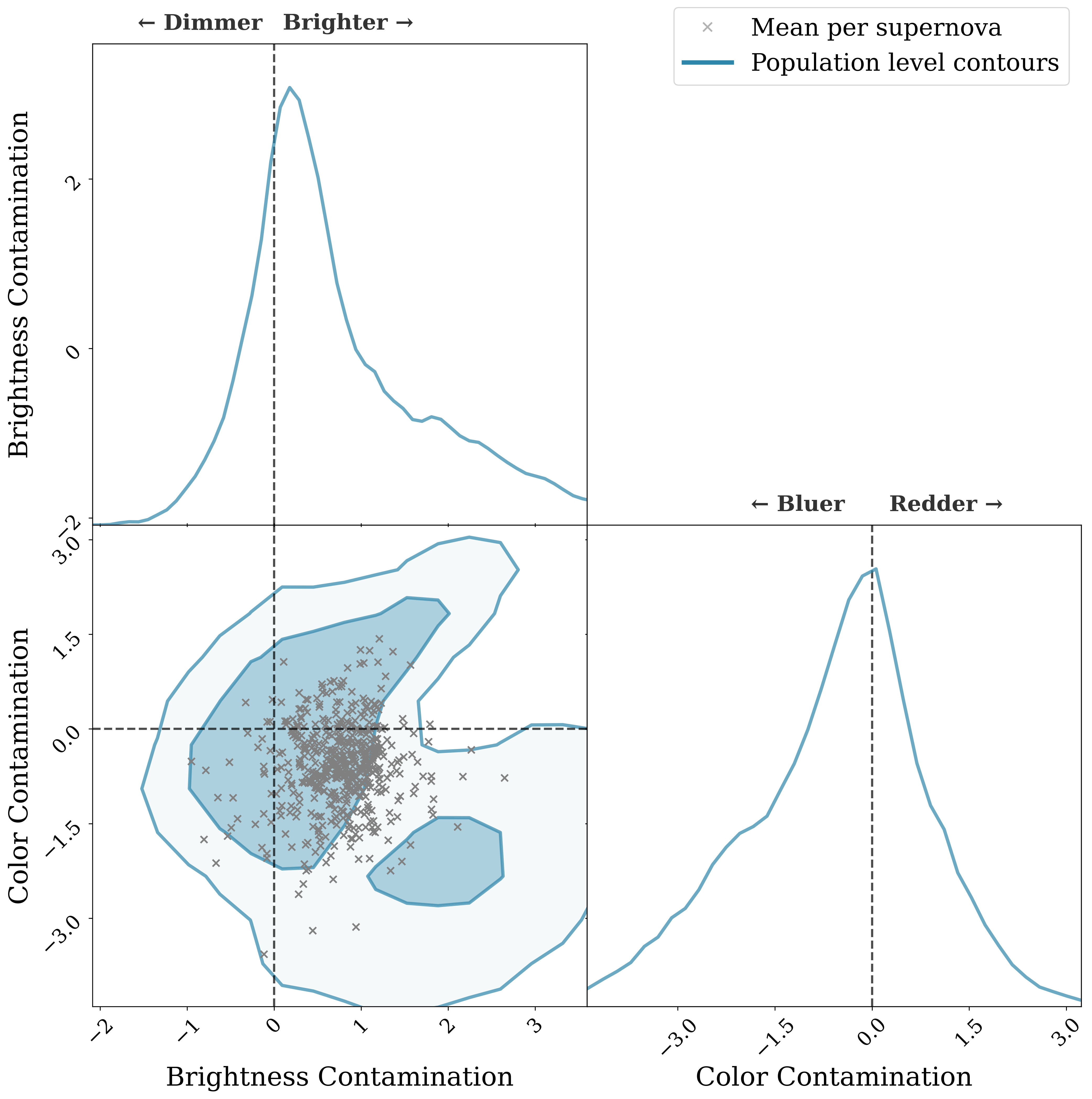}
\caption{Distribution of contamination metrics across the supernova sample. The diagonal panels show 1D marginal distributions for brightness contamination $C_{\mathrm{bright}}$ (top left) and colour contamination $C_{\mathrm{colour}}$ (bottom right). The bottom left panel shows the 2D joint distribution with 1$\sigma$ (inner) and 2$\sigma$ (outer) contours. Grey crosses mark individual supernova measurements. The positive mean brightness contamination and negative mean colour contamination indicate systematic biases in the anomalous data.}
\label{fig:contamination_triangle}
\end{figure*}

Figure~\ref{fig:contamination_triangle} presents the distribution of contamination metrics across the supernovae analysed. The brightness contamination $C_{\mathrm{bright}}$ showed a posterior mean of 0.718 $\pm$ 0.862 across the population, with a 1$\sigma$ confidence interval of [-0.065, 1.660]. Importantly, this contamination metric incorporates the probabilistic nature of anomaly detection: each data point contributes to the contamination score weighted by its probability of being anomalous, rather than through binary classification. The positive mean indicates that the net systematic effect of probable anomalies is to push flux measurements upward. Across all posterior samples, 63.8 per cent exhibit positive brightness contamination; this does not mean that 63.8 per cent of anomalies are bright, but rather that 63.8 per cent of the time, the probability-weighted contribution of all potentially anomalous points results in a net brightening effect. While this bias is less than 1$\sigma$ from zero and thus not definitively established, it suggests a systematic trend that, if left uncorrected in traditional analyses, could lead to underestimation of peak luminosity or overestimation of host galaxy extinction.

The colour contamination metric $C_{\mathrm{colour}}$ exhibited a posterior mean of -0.563 $\pm$ 0.989 across the population, with a 1$\sigma$ confidence interval of [-1.771, 0.207]. Like the brightness metric, this represents the probability-weighted net effect across all data points. The negative mean reveals that anomalous points preferentially affect blue bandpasses more than red ones. The posterior distribution shows 43.9 per cent of samples with negative colour contamination (where the probability-weighted effect pushes measurements bluer) versus 14.0 per cent with positive colour contamination (pushing redder), with the remainder showing negligible colour bias. Again, while the mean contamination is less than 1$\sigma$ from zero, the consistent direction of the bias across a substantial fraction of the posterior samples indicates a systematic tendency worth accounting for in precision cosmology. This blue preference in anomalous data provides insight into the systematic shifts observed in the colour parameter during the similarity analysis. When blue bands are preferentially contaminated with brighter measurements, traditional fitting methods that do not account for this contamination will infer artificially bluer supernovae (lower $c$ values), underestimating the true extinction. Through the Tripp relation ($\mu = m_B - M + \alpha x_1 - \beta c$), these systematically lower $c$ values would lead to underestimated distance moduli, making supernovae appear closer than they are and potentially biasing measurements of dark energy parameters and the Hubble constant.

The significant variation in both metrics, clearly visible in the broad posterior distributions in Figure~\ref{fig:contamination_triangle}, indicates substantial heterogeneity across the supernova sample. This heterogeneity demonstrates that contamination effects are not uniform across all supernovae but vary substantially from object to object. Some supernovae show strong blue contamination while others show red contamination, emphasising that our framework's ability to quantify these effects individually for each supernova is crucial. A blanket correction approach would fail to capture this diversity and could introduce new biases.

The contamination analysis provides quantitative evidence that the systematic differences between our anomaly detection framework and traditional methods, particularly the 0.418$\sigma$ average difference in the colour parameter, arise from the proper treatment of wavelength dependent contamination. By identifying and downweighting anomalous measurements that are preferentially blue and bright, our framework recovers more accurate colour parameters that better reflect the true supernova properties. This improved accuracy in individual supernova parameters will propagate through to more reliable cosmological constraints when these measurements are used for distance determinations in future analyses.

\subsection{Demonstrating the Framework: Case Studies}

Having established the global statistical performance, we now use three representative supernovae to demonstrate how the framework addresses different challenges in photometric data processing. The cases show: standard contamination mitigation via isolated outlier flagging (SN 21lnf, Section~\ref{subsec:sn21lnf}); automatic filter removal for systematically corrupted bandpasses (SN 21hwx, Section~\ref{subsec:sn21hwx}); and filter preservation, where only specific anomalous points are flagged within a bandpass (SN 19ekb, Section~\ref{subsec:sn19ekb}). For each case, we present both posterior distributions and light curve fits to compare our method with approaches that do not use automated anomaly detection.

\subsection{SN 19ekb: Filter Preservation and Tighter Constraints}
\label{subsec:sn19ekb}

\begin{figure*}
\centering
\includegraphics[width=0.8\textwidth]{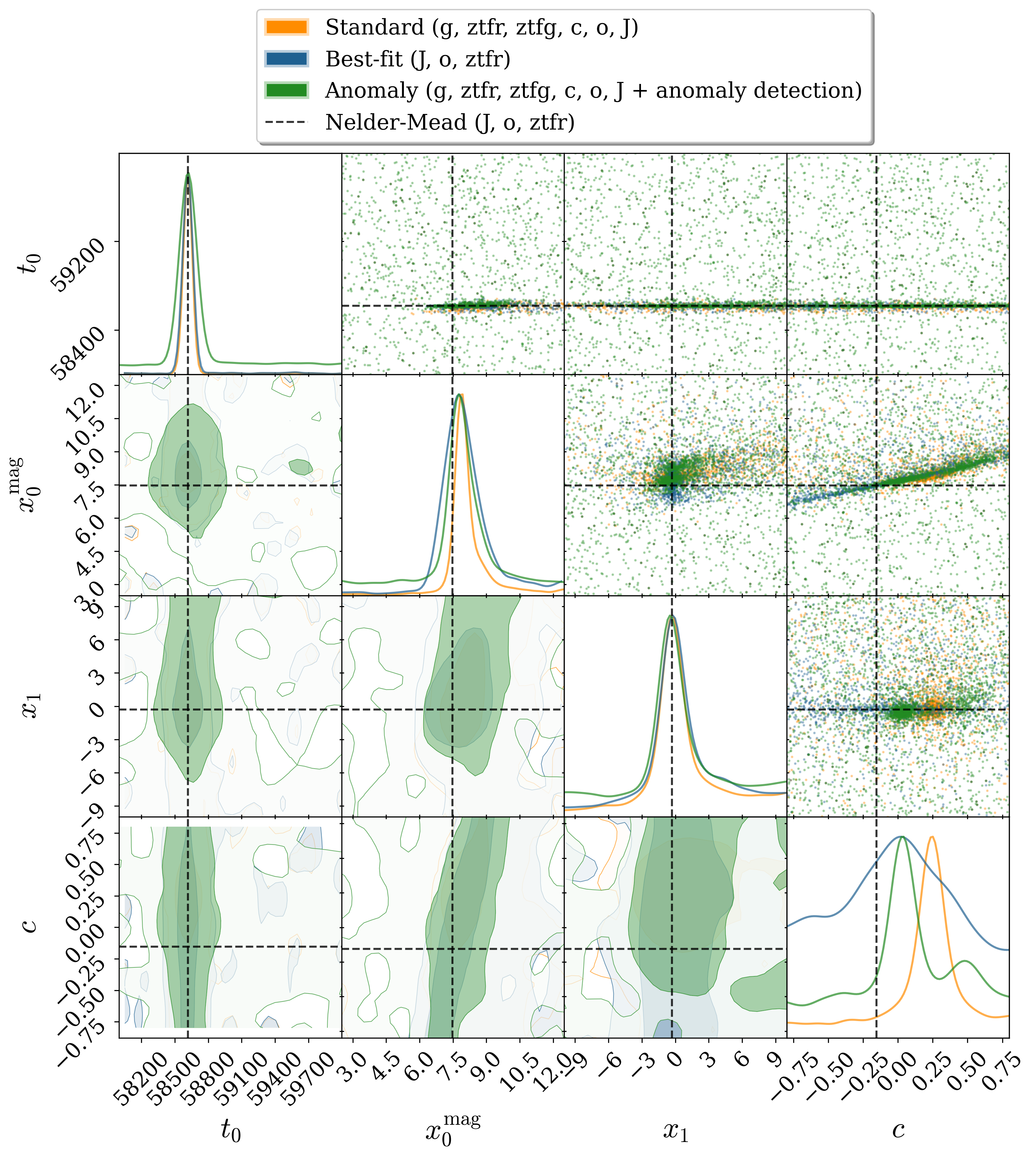}
\caption{Corner plot showing posterior distributions for SALT parameters $x_0$, $x_1$, and $c$ for SN 19ekb. Blue contours show the naive fit using all data (Case A), orange shows the systematic filter selection with removed filters (Case B), and green shows our Bayesian anomaly detection results (Case C). The anomaly detection achieves posteriors with tighter constraints by selectively removing outliers while preserving valid data.}
\label{fig:19ekb_corner}
\end{figure*}

\begin{figure*}
\centering
\includegraphics[width=\textwidth]{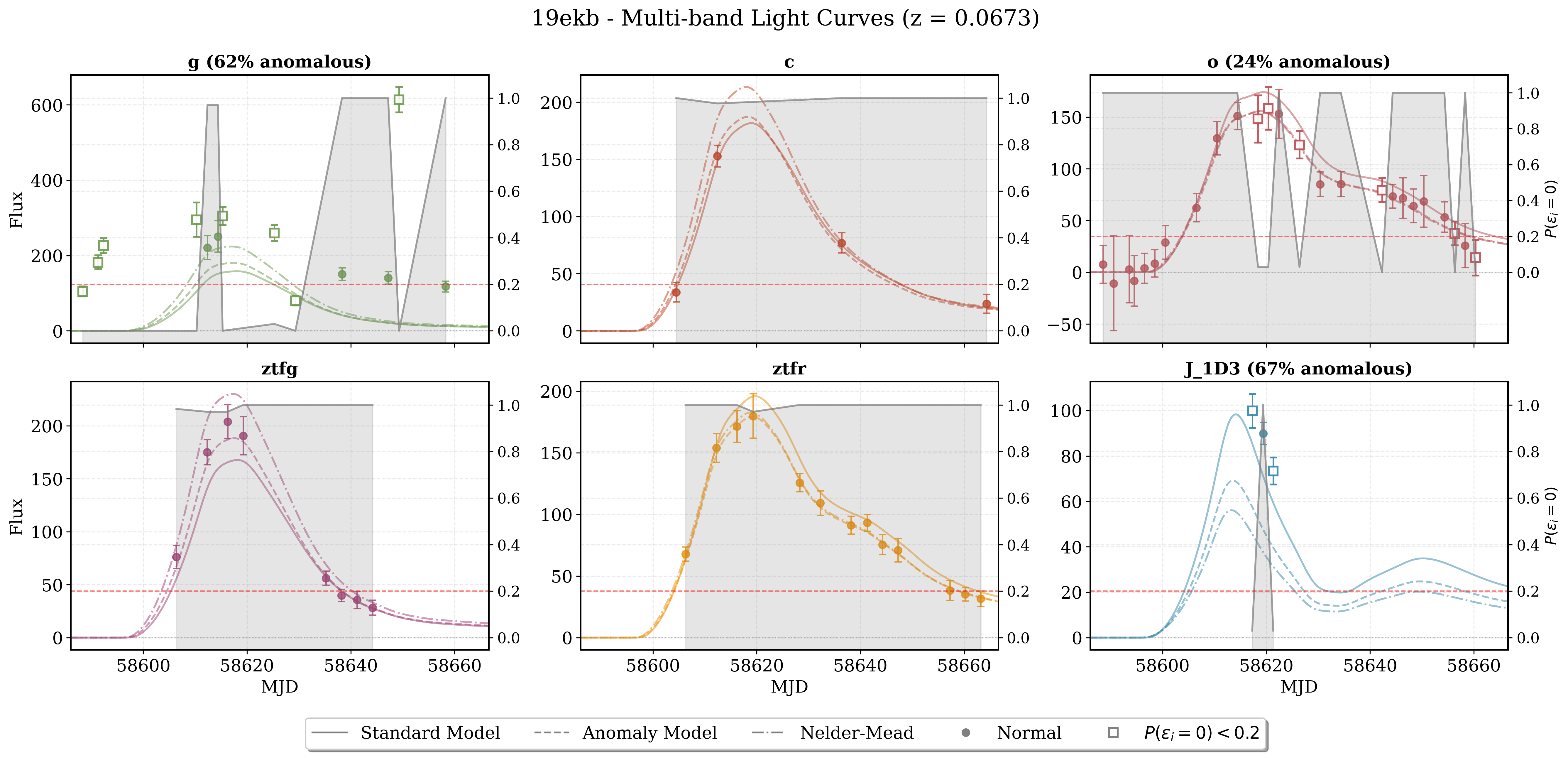}
\caption{Light curve fits for SN 19ekb, with each panel showing a different transmission filter fit. All filters were fit jointly, then split up for visualisation only. Points that were likely to be anomalous (i.e., $P(\varepsilon=0) < 0.2$) are marked with squares. Other points are marked with dots. The grey line and shading represent the probability that each data point was anomalous, with the scale shown on the right-hand axis. The ZTF g band (green), which is ignored by traditional analysis, is partially preserved here.}
\label{fig:19ekb_lightcurves}
\end{figure*}

The analysis of SN 19ekb illustrates the effect of filter preservation on parameter constraints. Figures~\ref{fig:19ekb_corner} and \ref{fig:19ekb_lightcurves} compare the posterior distributions for the SALT parameters from the three analysis cases. The naive fit (Case A; Section~\ref{subsec:analysis_cases}), which uses all data, yields tightly constrained but systematically biased posteriors, particularly for the colour parameter, $c$. The systematic filter selection approach (Case B; Section~\ref{subsec:analysis_cases}), in which the entire ZTF g and ATLAS c bands were removed based on our $\chi^2$ criterion, produces posteriors that are robust to this contamination, albeit with broader constraints.

In the anomaly detection framework (Case C; Section~\ref{subsec:analysis_cases}), the model automatically identifies and flags only the few anomalous points within the ZTF g and ATLAS c bands. By preserving the majority of valid data in these filters, Case C recovers parameter means consistent with the systematically cleaned data of Case B while achieving demonstrably tighter posterior constraints. This result shows that by retaining additional valid data, the framework can produce more precise parameter constraints compared to an analysis where entire filters are removed. This demonstrates that by retaining statistically valid data, our framework can improve the precision of supernova standardisation, which in turn could lead to tighter constraints on cosmological parameters that are crucial for addressing discrepancies such as the Hubble tension.

\subsection{SN 21hwx: Automatic Filter Removal and Identical Constraints}
\label{subsec:sn21hwx}

\begin{figure*}
\centering
\includegraphics[width=\textwidth]{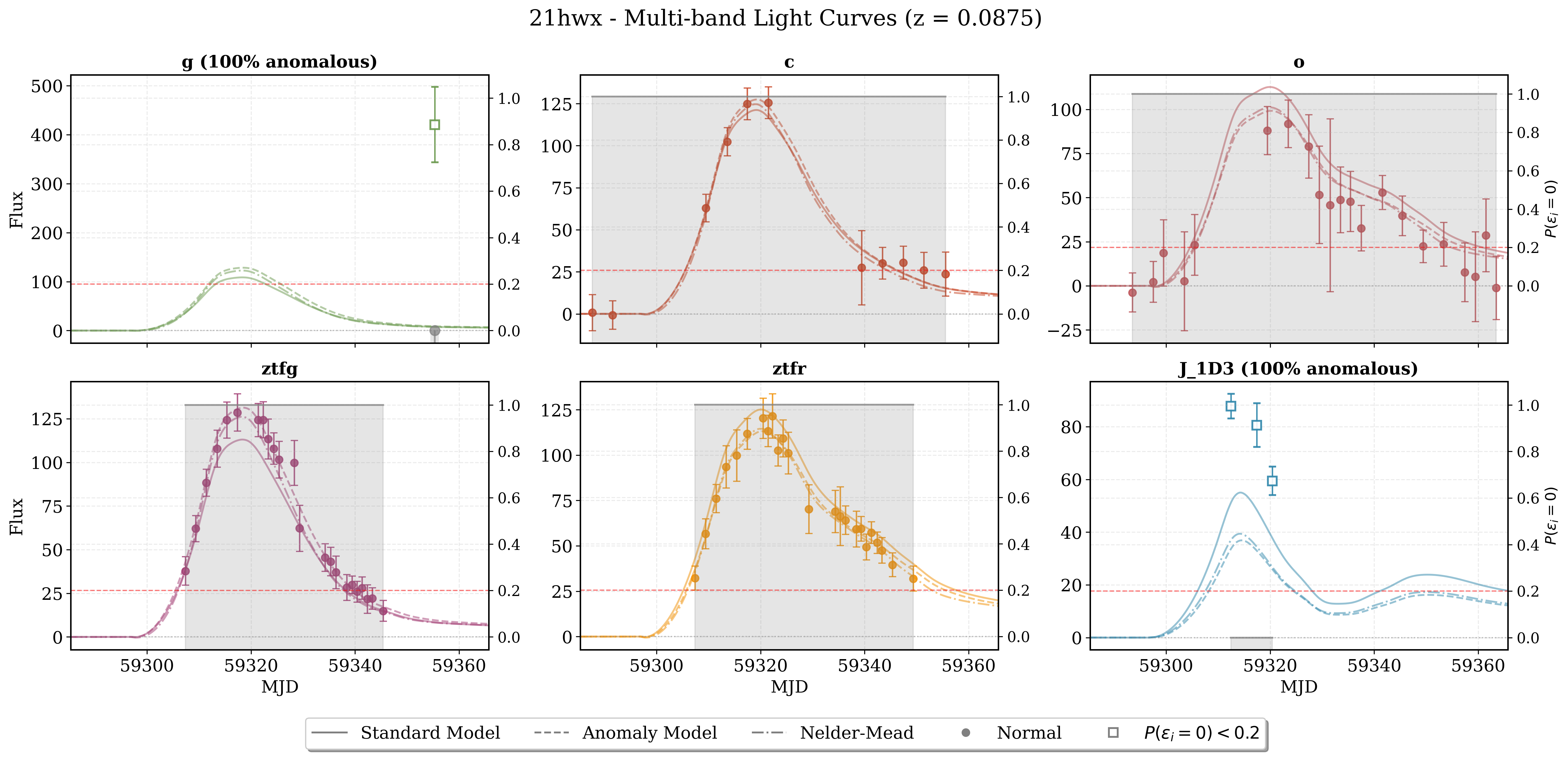}
\caption{Light curve fits for SN 21hwx showing the anomaly detection framework automatically flags all points in the problematic ZTF g and j bands, matching the systematic removal of these filters in the traditional method. The grey line and shading represent the probability that each data point was anomalous, with the scale shown on the right-hand axis.}
\label{fig:21hwx_lightcurves}
\end{figure*}

SN 21hwx presents a scenario where entire filters are systematically corrupted, demonstrating how the automated framework's decision aligns with manual analysis. The light curve in Figure~\ref{fig:21hwx_lightcurves} shows that the ZTF r, g, and j band data are entirely inconsistent with the supernova model.

\begin{figure*}
\centering
\includegraphics[width=0.8\textwidth]{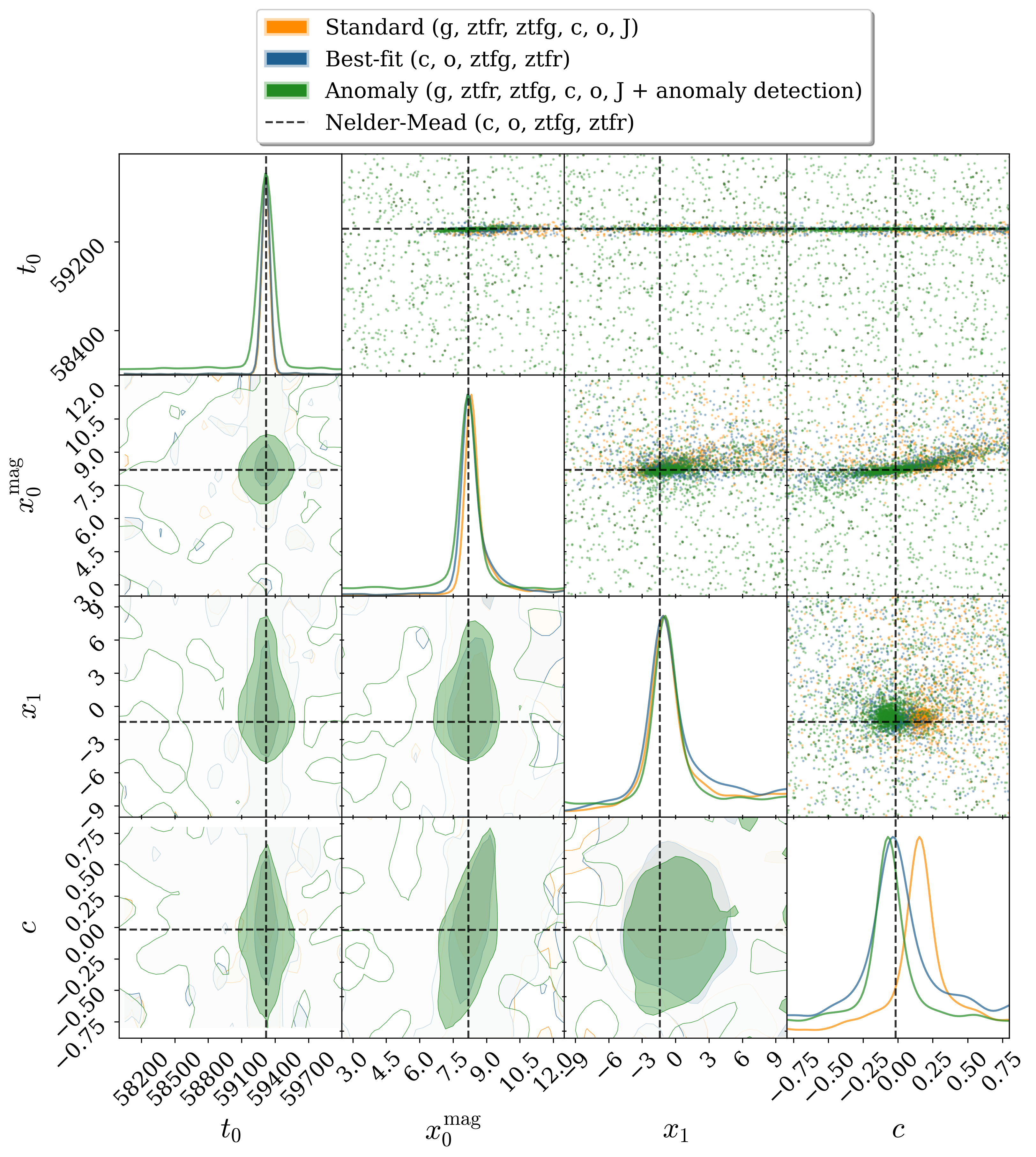}
\caption{Corner plot showing posterior distributions for SN 21hwx across the three analysis methods (Cases A, B, and C; Section~\ref{subsec:analysis_cases}). The anomaly detection framework automatically identifies that the entire ZTF g and j bands require removal, matching the outcome from our systematic filter selection and yielding identical posterior constraints. The bands noted in each legend indicate the filters that were used in the fit.}
\label{fig:21hwx_corner}
\end{figure*}

The naive fit (Case A; Section~\ref{subsec:analysis_cases}) is strongly biased by these corrupted filters, producing skewed and unreliable parameter estimates. In our systematic filter selection (Case B; Section~\ref{subsec:analysis_cases}), these filters were identified and removed based on the $\chi^2$ criterion. Our framework (Case C; Section~\ref{subsec:analysis_cases}) arrives at the same conclusion automatically, flagging every data point in the ZTF r, g, and j bands as anomalous. Because both the systematic filter selection and anomaly detection methods effectively remove the same problematic data in this case, they produce statistically identical posterior constraints that are robust to the contaminating data, as shown in Figure~\ref{fig:21hwx_corner}. This case demonstrates the framework's capacity to automate this important data quality decision in a statistically principled and reproducible manner.

\subsection{SN 21lnf: Selective Outlier Mitigation}
\label{subsec:sn21lnf}

\begin{figure*}
\centering
\includegraphics[width=\textwidth]{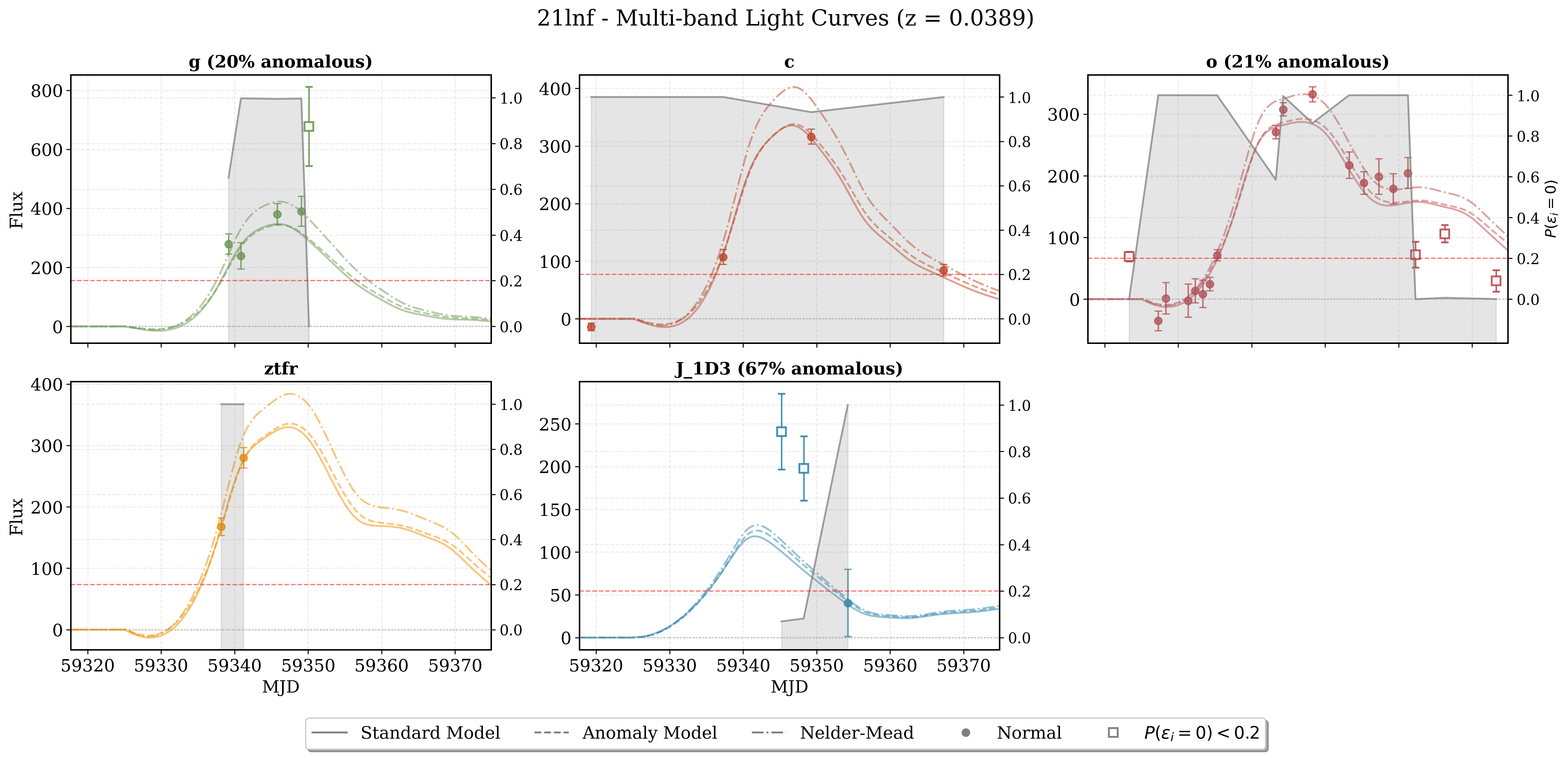}
\caption{Light curve fits for SN 21lnf showing selective flagging of individual outliers across multiple bands. The anomaly detection identifies specific problematic measurements marked as squares while retaining all valid data marked as dots. The grey line and shading represent the probability that each data point was anomalous, with the scale shown on the right-hand axis.}
\label{fig:21lnf_lightcurves}
\end{figure*}

The final example, SN 21lnf, represents a case with sporadic outliers scattered across multiple filters. As shown in the light curve plot (Figure~\ref{fig:21lnf_lightcurves}), the framework identifies and flags these individual anomalous points while preserving the surrounding data. A naive fit (Case A; Section~\ref{subsec:analysis_cases}) would be subtly biased by these points, while traditional filter selection approaches (Case B; Section~\ref{subsec:analysis_cases}) typically operate at the filter level rather than identifying individual outliers.

\begin{figure*}
\centering
\includegraphics[width=0.8\textwidth]{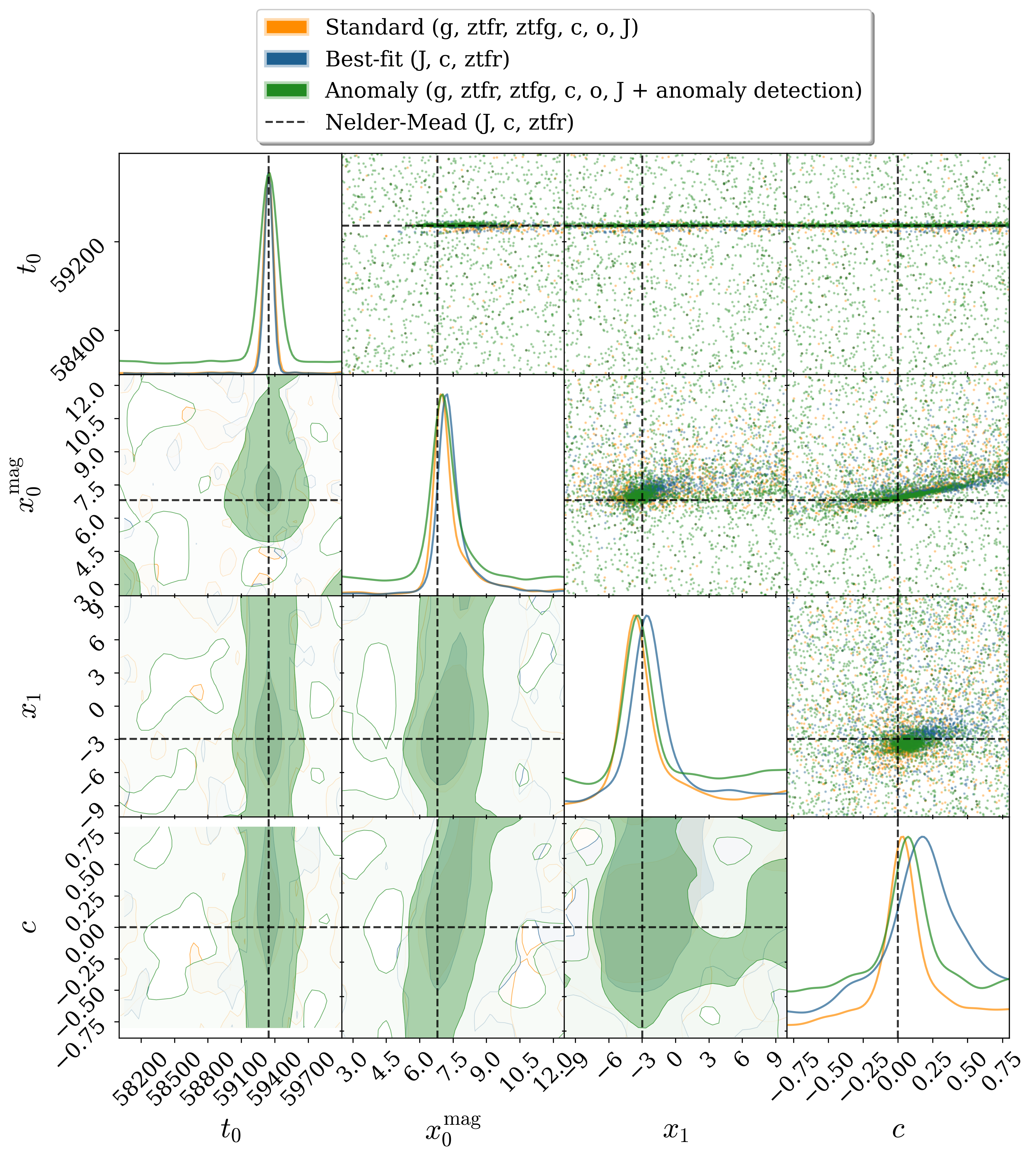}
\caption{Corner plot showing posterior distributions for SALT parameters for SN 21lnf across the three analysis methods (Cases A, B, and C; Section~\ref{subsec:analysis_cases}). The systematic filter selection and anomaly detection framework produce consistent parameter estimates, demonstrating the framework's ability to identify sporadic outliers.}
\label{fig:21lnf_corner}
\end{figure*}

The posterior distributions in Figure~\ref{fig:21lnf_corner} show close agreement between the systematic filter selection and our automated framework, indicating that both methods converge on a similar result free from the influence of obvious outliers. However, the objective flagging of several low-significance outliers by our framework results in a minor but systematic shift in the posterior for the colour parameter, $c$. This highlights the framework's ability to correct for low-level systematic errors introduced by contaminated data points that can be missed by traditional, filter-level analyses. The framework achieves this outcome through an automated process, which is a desirable property for ensuring consistency and scalability in large surveys.

\section{CONCLUSION}
\label{sec:conclusion}

In this work, we presented a Bayesian framework for the automated processing of Type Ia supernova light curves. This approach builds upon and extends the comprehensive data curation methods employed in the Hawaii Supernova Flows dataset, offering a fully automated alternative to manual inspection that becomes essential for future large scale surveys. We integrate an anomaly detection model directly into the likelihood function, treating the quality of each data point as a latent variable. This allows for simultaneous inference of the physical supernova model and probabilistic identification of data contamination, thereby reducing the reliance on manual preprocessing.

Our application to the high quality data from the Hawaii Supernova Flows survey demonstrated that this framework addresses several challenges in SNe Ia analysis. Firstly, it provides a mechanism for mitigating the influence of individual outliers on parameter estimates. Secondly, it automates filter selection by identifying systematically corrupted bandpasses in a reproducible manner. Thirdly, the framework facilitates improved data preservation by flagging specific anomalous epochs, which retains valid data points within a filter that might otherwise be discarded by conventional cuts. The statistical agreement (98.2 per cent of all SALT3 parameter estimates agreeing within their combined 1$\sigma$ uncertainty) demonstrates the reliability of our framework. 

Beyond identification, our contamination analysis quantified the systematic biases introduced by anomalous data. We found that anomalous points tend to be systematically brighter than model predictions ($C_{\mathrm{bright}}$ = 0.718 $\pm$ 0.862) and preferentially affect blue bandpasses ($C_{\mathrm{colour}}$ = -0.563 $\pm$ 0.989). If left uncorrected, these biases would lead to underestimated extinction and systematically lower distance moduli through the Tripp relation, potentially affecting dark energy constraints and Hubble constant measurements. These wavelength dependent contamination effects provide insight into the systematic differences observed in the colour parameter between our framework and traditional methods. The heterogeneous nature of these contamination effects across the sample validates our approach of object by object anomaly detection rather than blanket corrections.

This automated and statistical approach is well suited for the challenges of upcoming cosmological surveys. The data volume from the Vera C. Rubin Observatory's Legacy Survey of Space and Time will make manual inspection impractical, necessitating scalable and objective methods. Our framework offers one such pathway for processing large numbers of light curves, aiming to preserve the integrity and statistical power of the resulting cosmological samples, which will be essential for improving the precision of measurements that may help resolve the Hubble tension. While demonstrated for SNe Ia analysis with the SALT3 model, the underlying Bayesian methodology is general. It could be readily adapted for use with other SNe Ia light curve models, such as BayeSN \citep{Thorp2022} or SNooPy \citep{Burns2011}, or to other astrophysical domains where data may be affected by non Gaussian noise or contamination.

Ultimately, this work represents a crucial step towards building fully autonomous, end-to-end analysis pipelines for precision cosmology. Although demonstrated here with the SALT3 model, the framework's model-agnostic design makes it a versatile tool that can enhance any likelihood-based SNe Ia analysis, offering a pathway to more precise and reliable results across different modelling approaches. Subsequent work will assess the cosmological implications of these results by applying this refined dataset to a full cosmological parameter inference, quantifying how the removal of these systematic biases impacts key measurements such as the dark energy equation of state and the Hubble constant. By embedding data quality assessment within the core statistical model, we can move towards a more robust, efficient, and powerful era of cosmological discovery.

\section*{Acknowledgements}
This material is based upon work supported by the Google Cloud research credits program, with the award GCP397499138. This work was supported by the research environment and infrastructure of the Handley Lab at the University of Cambridge. SL is funded by the ERC through the UKRI guarantee scheme.

\section*{Data Availability}
The data underlying this article are available in the Hawaii Supernova Flows public data release \citep{do2025hawaii}. The code implementing the Bayesian anomaly detection framework is available at \url{https://github.com/samleeney/Bayesian-Anomaly-Detection}, with a permanent archive at \url{https://doi.org/10.5281/zenodo.17127169}, which provides several simple, easy to implement examples including the complete analysis pipeline used in this work.


\bibliographystyle{mnras}
\bibliography{references} % if your bibtex file is called references.bib

%%%%%%%%%%%%%%%%%%%%%%%%%%%%%%%%%%%%%%%%%%%%%%%%%%

%%%%%%%%%%%%%%%%% APPENDICES %%%%%%%%%%%%%%%%%%%%%

% \appendix
% \section{Appendix title}

%%%%%%%%%%%%%%%%%%%%%%%%%%%%%%%%%%%%%%%%%%%%%%%%%%

% Don't change these lines
\bsp    % typesetting comment
\label{lastpage}
\end{document}